\documentclass[twocolumn,prl,superscriptaddress,floatfix,showpacs]{revtex4-1}
\usepackage[utf8]{inputenc}
\usepackage{amsmath}
\usepackage{amssymb}
\usepackage{graphicx}

\usepackage{graphics}
\usepackage{epsfig}

\newcommand{\be}{\begin{equation}} \newcommand{\ee}{\end{equation}}
\newcommand{\bea}{\begin{eqnarray}} \newcommand{\eea}{\end{eqnarray}}

\begin{document}

\title{Self-trapping self-repelling random walks}

\author{Peter Grassberger}

\affiliation{JSC, FZ J\"ulich, D-52425 J\"ulich, Germany}

\date{\today}
\begin{abstract}
Although the title seems self-contradictory, it does not contain a misprint. The model we study is a 
seemingly minor modification of the ``true self-avoiding walk" (TSAW) model of Amit, Parisi, and 
Peliti in two dimensions. The walks in it are self-repelling up to a characteristic time $T^*$ (which 
depends on various parameters), but spontaneously (i.e., without changing any control parameter) become 
self-trapping after that. For free walks, $T^*$ is astronomically 
large, but on finite lattices the transition is easily observable. In the self-trapped regime, walks
are subdiffusive and intermittent, spending longer and longer times in small areas until they escape 
and move rapidly to a new area. In spite of this, these walks are extremely efficient in covering 
finite lattices, as measured by average cover times. 
\end{abstract}
\maketitle

Random walks are ubiquitous in nature, in science, and in technology. Be it the thermal motion of gas 
molecules \cite{Einstein}, the evolution of financial indices \cite{Bachelier,Bouchaud-Potters},
the foraging of an animal \cite{Benichou}, the Monte Carlo code of a scientist working in statistical
 physics \cite{Barkema}, the shape of a randomly coiled polymer in a good solvent \cite{deGennes}, 
or the carrying of a message in a random ad hoc network \cite{Avin}: They are all more or less 
described by random walks, and thus random walks have been among the most studied 
objects in mathematical statistics \cite{Spitzer}. But in most of these problems
they only represent a first crude approximation. In a gas or liquid, there is usually also convection. 
Financial time series show heavy tailed distributions \cite{Bouchaud-Potters}. And animal walks are 
not entirely random but also guided by the availability of food, and are often characterized by 
alternating periods of very slow and fast motion, what is often modelled as Levy flights \cite{levy}. 
One of the most common deviations from perfect randomness is that random walks often have memory.

Maybe the best studied model of walks with memory are self-avoiding walks (SAWs) \cite{Madras}, which 
describe the 
statistics of very long chain molecules, and where the ``memory" takes care of the fact that in a growing 
polymer, a new monomer cannot be placed onto a site that is occupied already. This 
modification implies that in less than 4 dimensions of space the characteristic size of a polymer made of 
$N$ monomers increases faster than $\sim N^{1/2}$. More precisely, the increase follows, for $d<4$, a power 
law $R\sim N^\nu$ with $\nu>1/2$, while $R/N^{1/2} \sim (\ln N)^\alpha$ with $\alpha = 1/4$ \cite{Clisby} 
at the upper critical dimension $d=4$. 

As pointed out by Amit {\it et al.} \cite{Amit}, while SAWs are indeed self-avoiding as geometrical objects,
they are as dynamical walks not self-{\it avoiding} but self-{\it killing}: When a walker tries to step on a 
site where she had already been, she is just killed. In what they called ``true self-avoiding walks" (TSAWs),
the walker instead tries to avoid in a short-sighted way to step on her own traces. Technically, this is 
implemented on a lattice by a walk where at each time step a unit of debris is dropped onto the site where 
the walker stands. As time goes on, a hilly landscape is formed where the height $h_i$ at site $i$ is just 
the amount of debris. The self-avoidance bias is then given by probabilities $p_j \propto e^{-\beta h_j}$ 
to step onto neighboring sites $j$, where $\beta$ plays the role of an inverse temperature. The 
self-avoidance is negligible for large temperature, while it is strongest for $\beta = \infty$. But even 
then its effect is much milder than in the original SAW model. No walker is killed, but they just try
to turn away gently. In the mathematical literature, such walks are often called ``self-repelling". 

In \cite{Amit} it was shown that the upper critical dimension for TSAWs is not $d=4$ but $d=2$. Thus they 
show trivial scaling for $d>2$, while they are swollen, $R\sim N^\nu$ with $\nu>1/2$, for $d=1$. For $d=2$ 
there should be again logarithmic corrections, but the exponent in the ansatz $R/N^{1/2} \sim 
(\ln N)^\alpha$ is not known, in spite of considerable efforts \cite{Amit,Obukhov,Derkachov}. 
A first attempt to obtain $\alpha$ was made in \cite{Amit}, where an effective field theory was proposed 
in which the bias of the walk was -- in a coarse-grained picture amenable to renormalization group (RG) 
ideas -- coupled 
to the average local slope of $h_i$. It was neglected that the walker is not only influenced by the 
gradient of the landscape, but also by its {\it roughness}. As shown by Obukhov and Peliti \cite{Obukhov}, 
this is not justified. It is well known that random walkers in rough landscapes are hindered by obstacles 
\cite{Bouchaud}, so roughness tends to make them move more slowly. The RG scheme proposed in \cite{Obukhov} 
was later criticized by \cite{Derkachov}, who pointed out that one has in general to consider 
also higher order couplings (beyond slope and roughness), which makes the problem non-renormalizable.
In \cite{SM} we argue that $\alpha=1/2$.

Apart from these formal problems, the scheme proposed in \cite{Obukhov} is also sick for a very basic
reason. In an RG treatment of TSAWs, one has to consider not only the RG flow, but also the flow of time.
Indeed, TSAWs are not stationary, and they are not even time reversal invariant \cite{SM}. As the landscape 
grows, its effect on the walker becomes stronger and stronger. 

To see this more quantitatively, let us consider TSAWs on a large but finite 
lattice of size $L\times L$. For convenience we take a square lattice with periodic 
(or, for easier coding, helical; the difference between them is negligible for the lattice 
sizes considered here) boundary conditions. The walker starts on a flat landscape $h_i=0$. If 
there were no self-repulsion (i.e. $\beta =0$), the lattice would be covered after a time 
$T_{\rm cover} \sim (4/\pi) L^2 (\ln L)^2$ \cite{Dembo,Grass-cover}. After that, the average 
height still grows linearly with time, but its roughness also grows without limits
\cite{Freund}. The variance of the height profile, 
\be
   \sigma(T) = L^{-2} \sum_i h_i^2 - T^2,
\ee
increases proportionally to $T$, and \cite{Freund}
\be
   \sigma(T)/T = \frac{4}{\pi} \ln L + {\cal O}(1) \quad {\rm for} \;\;\; L\to\infty.
\ee
For non-zero $\beta$, in contrast, it was conjectured \cite{Avin} that
\be
   T_{\rm cover} \sim a_T(\beta) L^2 \ln L .
\ee

\begin{figure}
\begin{centering}
\includegraphics[scale=0.30]{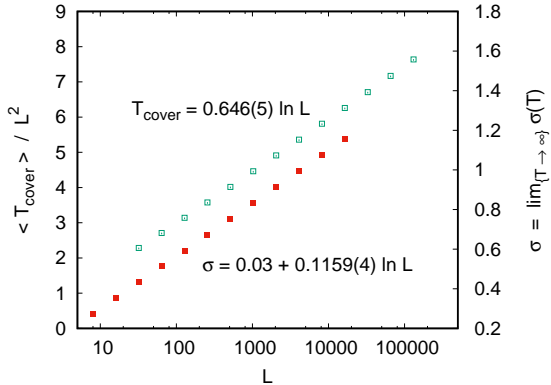}
\par\end{centering}
\caption{\label{fig1} (color online)  Log-linear plots of statistics of TSAWs with $\beta=\infty$
   on square lattices of size $L\times L$ with helical boundary conditions. The upper curve (open
   squares and left-hand y-axis) shows the average cover times, divided by $L^2$. The lower curve (filled
   squares and right-hand y-axis) shows the asymptotic height variances of the debris
   field. In both cases, error bars are much smaller than symbol sizes.}
\end{figure}

For $\beta=\infty$ this is indeed shown in Fig.~1, but completely analogous results were obtained also for 
finite $\beta$. The prefactor $a_T(\beta)$ diverges of course for $\beta\to 0$.

For the height variance for $T\gg T_{\rm cover}$, the effect of self-repulsion is even stronger. This 
time the variance stays finite for $T\to \infty$, with \cite{Freund}
\be
   \sigma(T) \sim a_\sigma(\beta) \ln L,       \label{varL}
\ee
see also Fig.~1 for $\beta = \infty$. Again the prefactor $a_\sigma(\beta)$ diverges as $\beta\to 0$.
From plots analogous to Fig.~1 (but for other values of $\beta$) we obtain
\be
   a_\sigma(\beta) \simeq 0.317(4)/\beta
\ee
for $\beta\to 0$.

In the RG treatment in \cite{Obukhov,Derkachov} it was assumed that one can start perturbatively 
around the point where both coupling constants (that for the slope and that for the roughness) are 
small. But as we have just seen, when the coupling to the slope is small, the roughness
increases for late times beyond any limit. Thus a perturbative treatment in the combined effects of 
roughness and slope becomes impossible.

In order to avoid this problem, one can change the model so that the landscape becomes less rough.
One possibility would be to let the debris diffuse. This could be presumably efficient, but it is 
rather awkward (and slow, from a numerical point of view) to implement -- and it is very likely that it 
will lead to problems similar to those discussed below. Much easier seems the following change: 
Instead of dropping all debris onto the site of the walker, only a fraction $1-\epsilon$ is dropped 
there. The rest is distributed uniformly among all of its neighbors.

\begin{figure}
\begin{centering}
\includegraphics[scale=0.30]{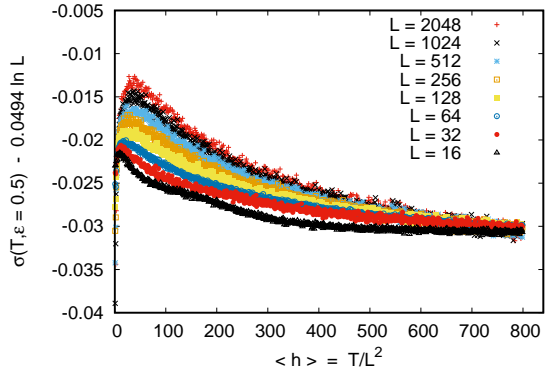}
\par\end{centering}
\caption{\label{fig2} (color online)  Plot of $\sigma(T) - a_\sigma(\beta=\infty,\epsilon) \ln L$ against the 
   average debris height $\langle h\rangle$, for $\epsilon=1/2$. According to Eq.~(\ref{varL}), these curves
   should all approach the same horizontal line for $T\to\infty$. The fact that they do this very slowly 
   and in a non-monotonic way seems to be a peculiarity of the $\beta=\infty$ limit on the square 
   lattice. It is neither seen for finite $\beta$ nor on the triangular lattice.}
\end{figure}

As seen from Fig.~2 for for $\beta=\infty$, this seems 
indeed to work -- at least on the square lattice and for $\epsilon = \epsilon_c = 1/2$. The variance 
increases still roughly according to Eq.~(\ref{varL}), but the prefactor -- called now 
$a_\sigma(\beta,\epsilon)$ -- is $\lesssim 0.05$. Indeed, Fig.~2 does not 
show $a_\sigma(\beta,\epsilon)$ or $\sigma(T)$, but rather $\sigma(T) - a_\sigma(\beta=\infty,\epsilon) \ln L$.
For reasons that are not fully understood, $\sigma(T)$ does not increase monotonically. This
anomaly seems to be related to the fact that walks have strongly reduced randomness for $\beta=\infty$.
It is even enhanced for $\epsilon < \epsilon_c$
\cite{SM}. For finite $\beta$ this anomaly is absent, and the asymptotic value of $\sigma(T)$ is reached 
monotonically. The latter is true also for the triangular lattice (with 
$\epsilon_c = 2/3$), and if debris on the square lattice is dropped not only onto the 4 nearest neighbors, 
but also (with the same amounts) onto the 4 next-nearest neighbors. In the last case we also found 
$\epsilon_c = 2/3$ \cite{SM}.

\begin{figure}
\begin{centering}
\includegraphics[scale=0.30]{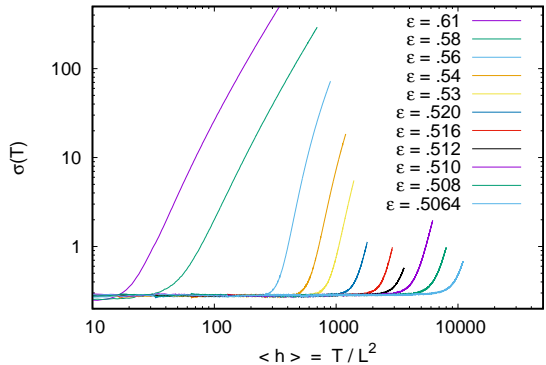}
\par\end{centering}
\caption{\label{fig3} (color online) Log-log plot of $\sigma(T)$ 
   against the average debris height $\langle h\rangle$, for square lattices with $L=512$. Each curve 
   corresponds to a different value of $\epsilon$. They are roughly horizontal up to a characteristic 
   debris height $h^*$ that increases roughly as an inverse power of $\epsilon = 1/2$, but deviations from
   such a power law are much larger than statistical errors.}
\end{figure}

For $\epsilon > \epsilon_c$ things change, however, completely. As seen in Fig.~3, $\sigma(T)$ first 
approaches rapidly a constant, but finally increases beyond limit as $T\to\infty$. The data in Fig.~3 are
for the square lattice with $L=512$ and $\beta=\infty$, but similar results were seen also in all other 
cases. In particular, nearly identical plots are obtained for $L=256$ and $L=1024$, the only difference 
being tiny shifts compensating the height differences of the curves before they start to rise. This 
means that the rise of $\sigma(T)$ starts
at a fixed debris height, not at a fixed time. This implies also that the same rise should also be seen
on an infinite lattice, because debris height increases also there with time. Since this increase is only
logarithmic on an infinite lattice, the transition happens there at astronomically large times, making 
it de facto unobservable.

\begin{figure}
\begin{centering}
\includegraphics[scale=0.30]{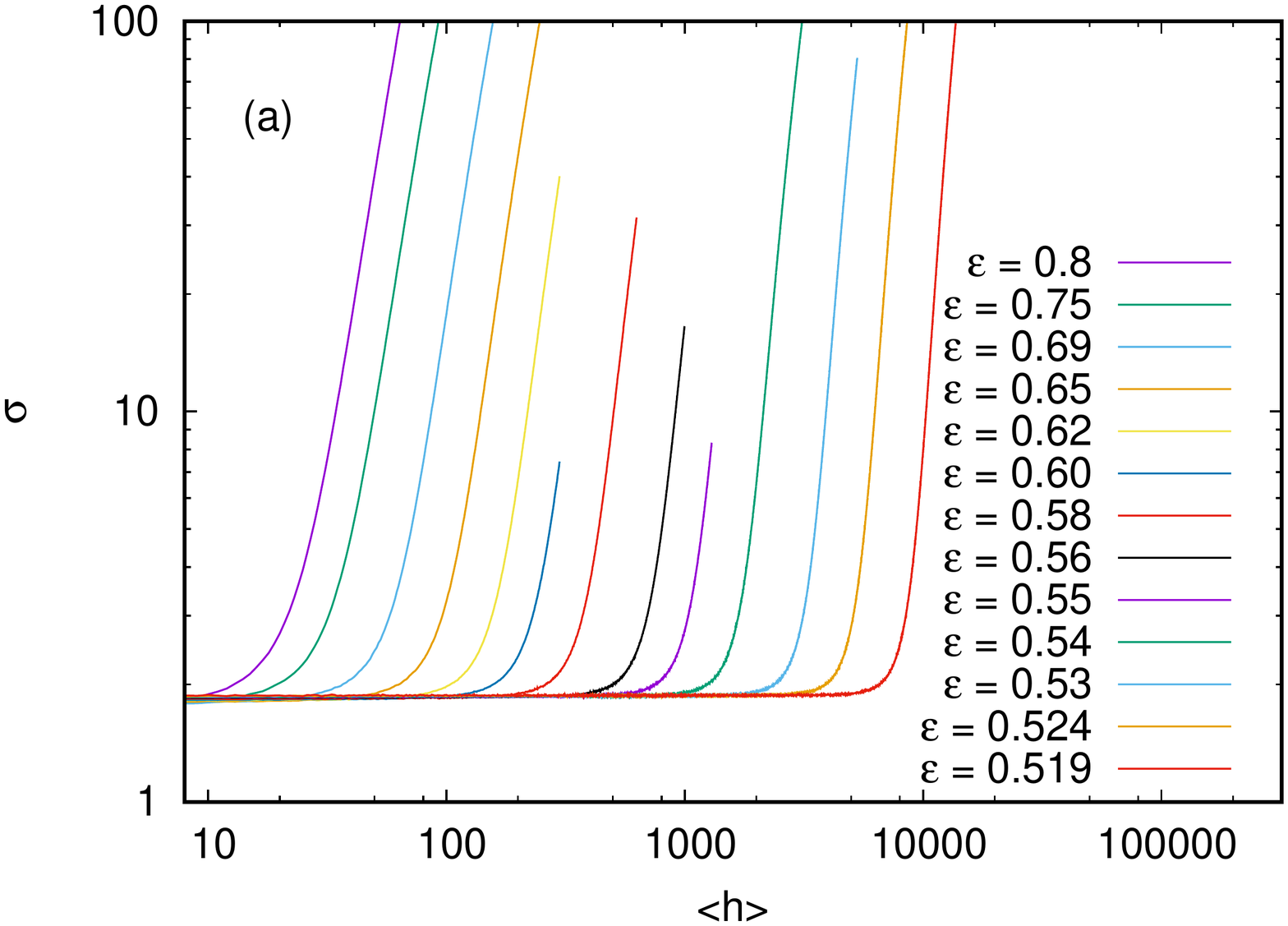}
\includegraphics[scale=0.30]{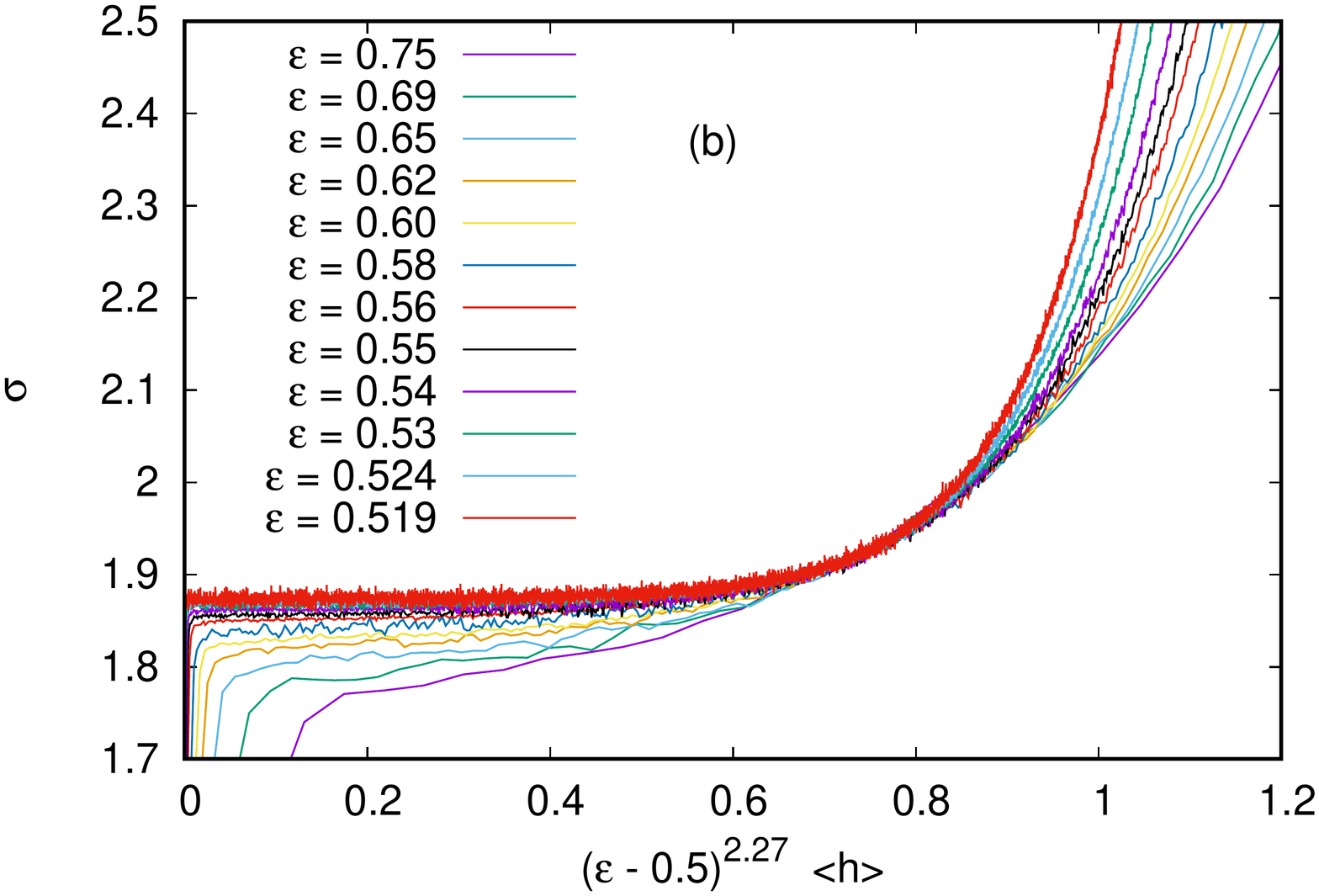}
\par\end{centering}
\caption{\label{fig4} (color online) (a) Plot of $\sigma(T)$ against  $\langle h\rangle$, but for $\beta=1.0$.
   (b) Same data, but plotted against $(\epsilon - 1/2)^{2.27} \langle h\rangle$.}
\end{figure}

Roughly, the characteristic densities $h^*$ in Fig.3 (at which roughness starts to increase) scale as 
$h^*\sim (\epsilon - 1/2)^{-2}$, but deviations from this are huge. The reason is most likely the same 
as that for the non-monotonicity in Fig.~2. Much more regular behavior is found for finite $\beta$ and on the 
triangular lattice. Results for $\beta=1$ on the square lattice are shown in Fig.~4. In panel (a) we show 
$\sigma$ versus $\langle h\rangle$, while the data are plotted against $(\epsilon - 1/2)^{2.27} \langle h\rangle$ 
in panel (b). The latter suggests strongly that (i) $\epsilon_c = 1/2$ is exact; (ii) The characteristic height 
scales as $h^* = c /(\epsilon - 1/2)^\gamma$ with $c=0.80(5)$ and $\gamma=2.27(2)$; and (iii) At $h=h^*$, the 
rise of $\sigma$ against $(\epsilon - 1/2)^\gamma h$ becomes infinitely steep for $\epsilon\to 1/2$. 

\begin{figure}
\begin{centering}
\includegraphics[scale=0.30]{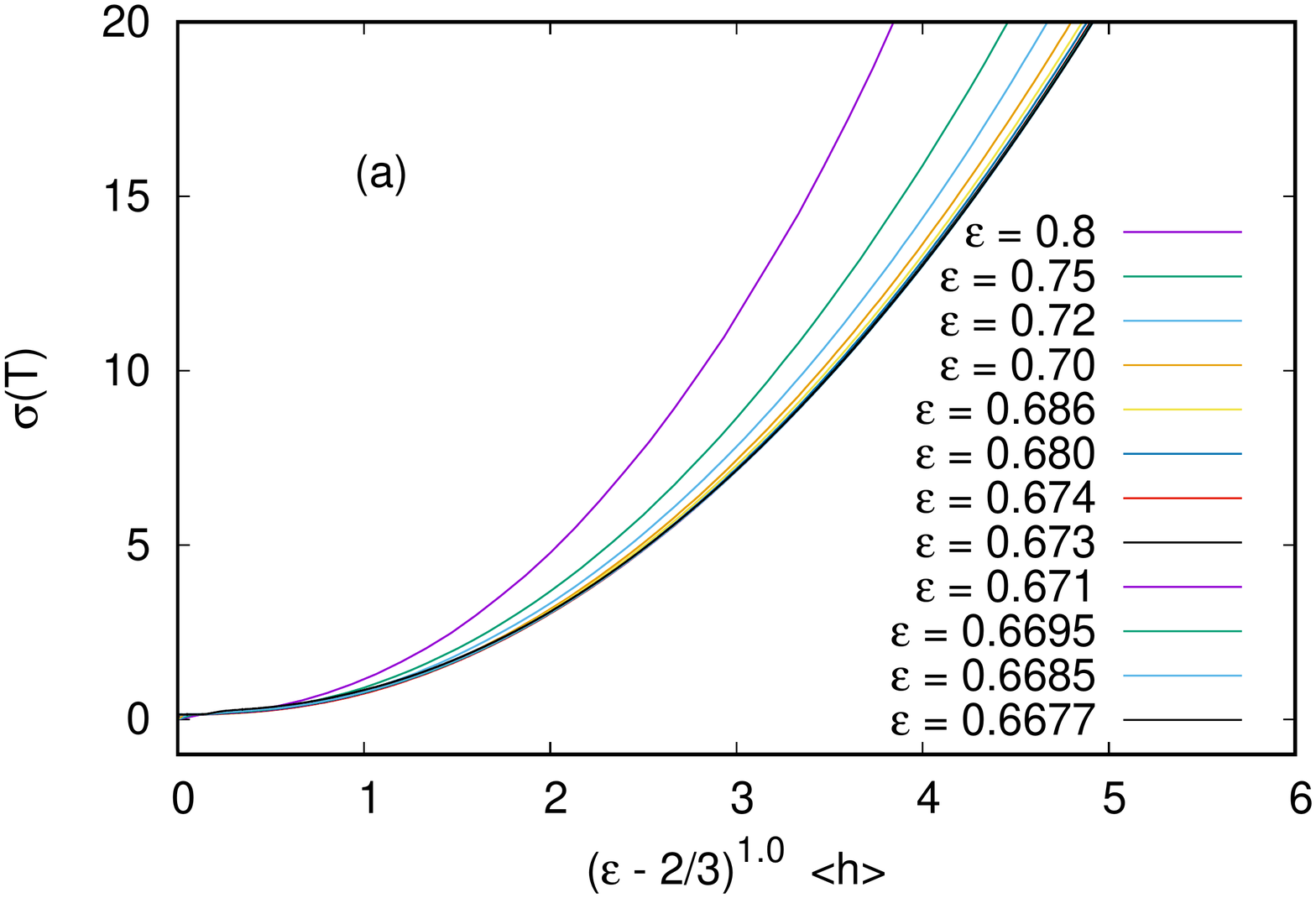}
\includegraphics[scale=0.30]{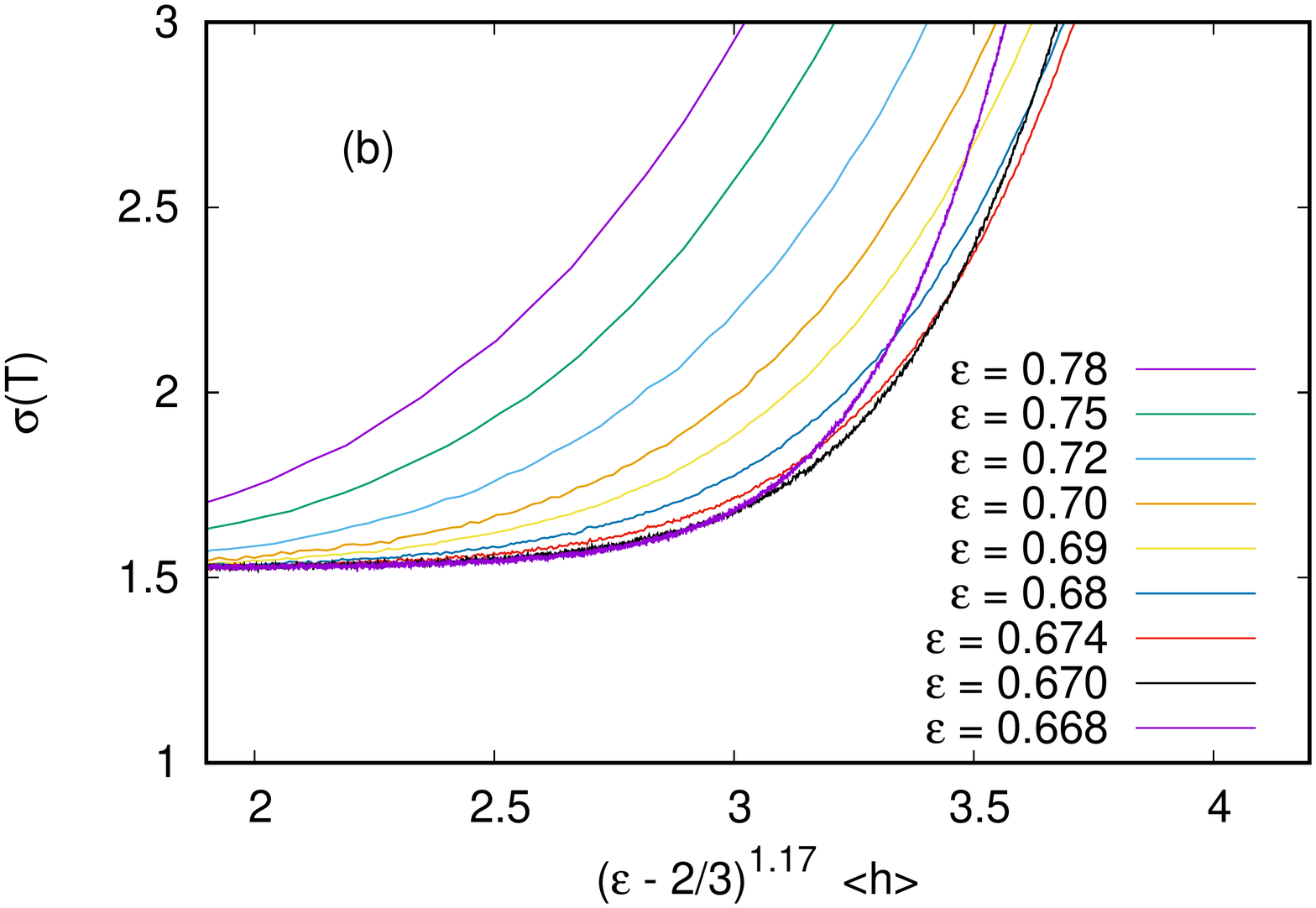}
\par\end{centering}
\caption{\label{fig5} (color online) Plots of $\sigma(T)$ against  $(\epsilon - 1/2)^\gamma \langle h\rangle$,
   for triangular lattices. Panel (a) is for $\beta=\infty$ and $\gamma = 1.0$, while (b) is for $\beta=1.0$
   and $\gamma = 1.17$.}
\end{figure}

Basically the same results were found also for $\beta=0.2$ and $\beta=5.0$. In particular, also there 
$\epsilon_c$ seems to be exactly $1/2$ and the same scaling seems to hold for $h^*$, with $c=4.3(5)$ 
for $\beta=0.2$ and $c=0.26(3)$ for $\beta=5.0$. The values for the exponent $\gamma$ are $2.24(3)$ 
and $2.26(3)$.

This suggests that $\gamma$ is universal, but this is shattered by the results for the triangular 
lattice. There, $\epsilon_c = 2/3$ (again for all values of $\beta$), but plots analogous to Fig.~4b
for $\beta=\infty$ (see Fig.~5a) and $\beta=1$ (see Fig.~5b) indicate that $\gamma\approx 1$ in
both cases More precisely, for $\beta=\infty$ we obtained $\gamma=1.00(2)$, while $\gamma \leq 1.17$ for 
$\beta=1$ (a more precise estimate for the latter is prevented by large corrections to scaling). 
Finally, we simulated also walks on the square lattice where the four next-nearest neighbors received 
the same amount of debris as the four nearest neighbors. The data \cite{SM} gave again $\epsilon_c = 
2/3$ for all $\beta$ and $\gamma = 1.00(1)$ for $\beta=\infty$, while the estimate $\gamma \leq 1.15$ 
for $\beta=1$ is again
affected by large corrections to scaling. In summary, it seems that there are two distinct universality 
classes, one with $\gamma\approx 1$, and the other with $\gamma \approx 2$. Within each class, there 
are still minor but statistically significant differences. The origin 
of this is not clear.

\begin{figure}
\begin{centering}
\includegraphics[scale=0.30]{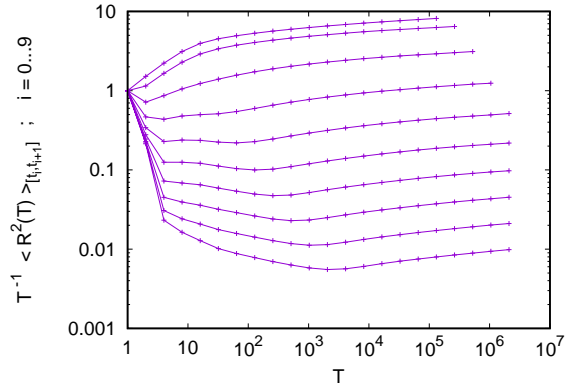}
\par\end{centering}
\caption{\label{fig6} (color online) Squared average end-to-end distances (divided by $T$) of the $T$ last 
   steps of walks with total length $t \in [t_i,t_{i+1}],\;\; 0\leq i<10,\; t_0=0,\; t_{i+1} = 2t_i +4L^2$.
   The top curve is for $i=0$, the bottom one for $i=9$. Parameters are $L=16384$, $\beta=\infty$ and 
   $\epsilon=0.7$.}
\end{figure}

For $h >h^*$, walks are subdiffusive and get more and more so as $h$ increases further. Let us define 
the average squared end-to-end distance of the last $T$ steps of a walk of total length $t$, averaged over 
$t\in [t_a,t_b]$, as $\langle R^2(T)\rangle_{[t_a,t_b]}$.
In Fig.~6 are plotted $T^{-1} \langle R^2(T)\rangle_{[t_i,t_{i+1}]}$ for $t_{i+1} = 2t_i +4L^2$, with $t_0=0,
0\leq i<10$, and $L=16384$. We see that the walks are stretched for all $T$ for $i=0,1$, and remain stretched
for large $T$ even when $i=2$ or 3. But for larger $i$ we see $R^2 <T$, mainly because the walks are strongly 
compressed for very small $T$. 

Thus, most of the time the walks are confined to narrow regions for short intervals 
whose length increases with $i$, while the evolution on larger time scales is characterized by
escape from these regions. Obviously, a typical walk stays for some time trapped in a region where $h$ was 
originally lower than average. As time goes on, it fills up the debris in this region, but it also builds 
a wall around it. When finally $h$ is so large that the walk escapes, it has built such a 
high wall that it gets trapped even longer in a neighboring region, etc. This scenario
is supported by the entropy of the walks, which is just equal to the entropy
provided by the random number generator. Entropies decrease fast (roughly exponentially)
with $\langle h\rangle$ \cite{SM}, implying that for large times the walk is hardly random at all.

We have seen that self-repelling walks become self-trapping when the debris height increases above
a critical height, if sufficiently much of the debris is placed on neighboring sites. The critical height
depends on this amount and on the type of lattice, but it is independent of the size of the lattice. 
Since the average debris height increases also for infinite lattices, this transition should be also seen 
there. Since this increase is however very slow ($\sim \ln T$), the self-trapping transition on infinite 
lattices should
be seen only at extremely large times, much larger than what is reachable with present-day computers.
Therefore, also lattice covering times should not -- at presently reachable values of $L$ -- be affected
by self-trapping, unless $\epsilon$ is extremely large. For the square lattice with $\epsilon=0.8$, e.g.,
we found that Eq.~(3) holds for $\beta=\infty$ with $\sigma_T(\infty) = 0.024(2)$. Thus walks with 
large $\epsilon$ should be optimal for disseminating/collecting information on large systems
(notice that our results should also apply on geometric random graphs \cite{Avin}).
Even faster could be walks where also next-nearest neighbors of visited sites are marked, 
but then the increased efficiency in terms of number of steps should be balanced against increased effort in
marking these sites.
% This holds, e.g., also for claims that Levy walks are optimal for harvesting \cite{Stanley}, 
% when the increased effort to make longer jumps is not taken into account.

%More generally, 
%Our results should also apply when the lattice is replaced by geometric random graphs.
%There, of course, the self-repulsion should not depend on the absolute number of previous visits to a node,
%but on the number of visits divided by its degree \cite{Avin}. More generally, our results can be 
%seen as a case where too much greed leads to a break down of an algorithm -- although
%the break down does not occur in applications that one typically might be interested in.

%\begin{figure}
%\begin{centering}
%\includegraphics[scale=0.30]{covertimes.pdf}
%\par\end{centering}
%\caption{\label{fig1} (color online)  Log-linear plot of average cover time for 2-toruses of size 
%  that would, however, leave the cover times monotonically decreasing with $L$.}
%\end{figure}

I am indebted to Gerhard Gompper, Dmitry Fedosov, and Sandipan Mohanty for most useful discussions.

\end{document}

\title{Supplemental material for ``Self-trapping self-repelling random walks"}

\author{Peter Grassberger}

\maketitle

\section{Irreversibility of TSAW's}

It is easily seen for simple short true self-avoiding walks that their probability is in general 
not symmetric under time reversal. Take, e.g., the 4-step walk with bends left-left-straight. 
In the TSAW ensemble for $\beta=\infty$ it has weight 
$\frac{1}{4}\times\frac{1}{3}\times\frac{1}{3}\times\frac{1}{2}$,
but the time-reversed walk with bonds straight-right-right has weight
$\frac{1}{4}\times\frac{1}{3}\times\frac{1}{3}\times\frac{1}{3}$. 

A priori, this asymmetry might be important only for short walks, but this is not true. In 
Fig.~S1 we plotted the average debris height at the position $x(t)$ of walks of total length $T$
against $t/T$. We see that: 
\begin{itemize}
\item This increases as $h \sim const + a \ln T$ with $a=0.036(1)$. Qualitatively, this is as for 
ordinary random walks, but the increase there is faster by roughly one order of magnitude;
\item Secondly, the very first and last parts of the walks visit sites are revisited less often than 
average. Again this is qualitatively as for ordinary random walks;
\item Finally, the curves are clearly asymmetric under $t \to T-t$.
At later times, the walker visits more often sites with large $h$. This asymmetry -- which is absent
for ordinary random walks -- is not decreasing with $T$. Instead, $\langle h(t)\rangle \sim 
a \ln T f(t/T)$ for large $T$ with a asymmetric scaling function $f(.)$.
\end{itemize}
 
\section{End-to-end distances}

\begin{figure}
\begin{centering}
\includegraphics[scale=0.30]{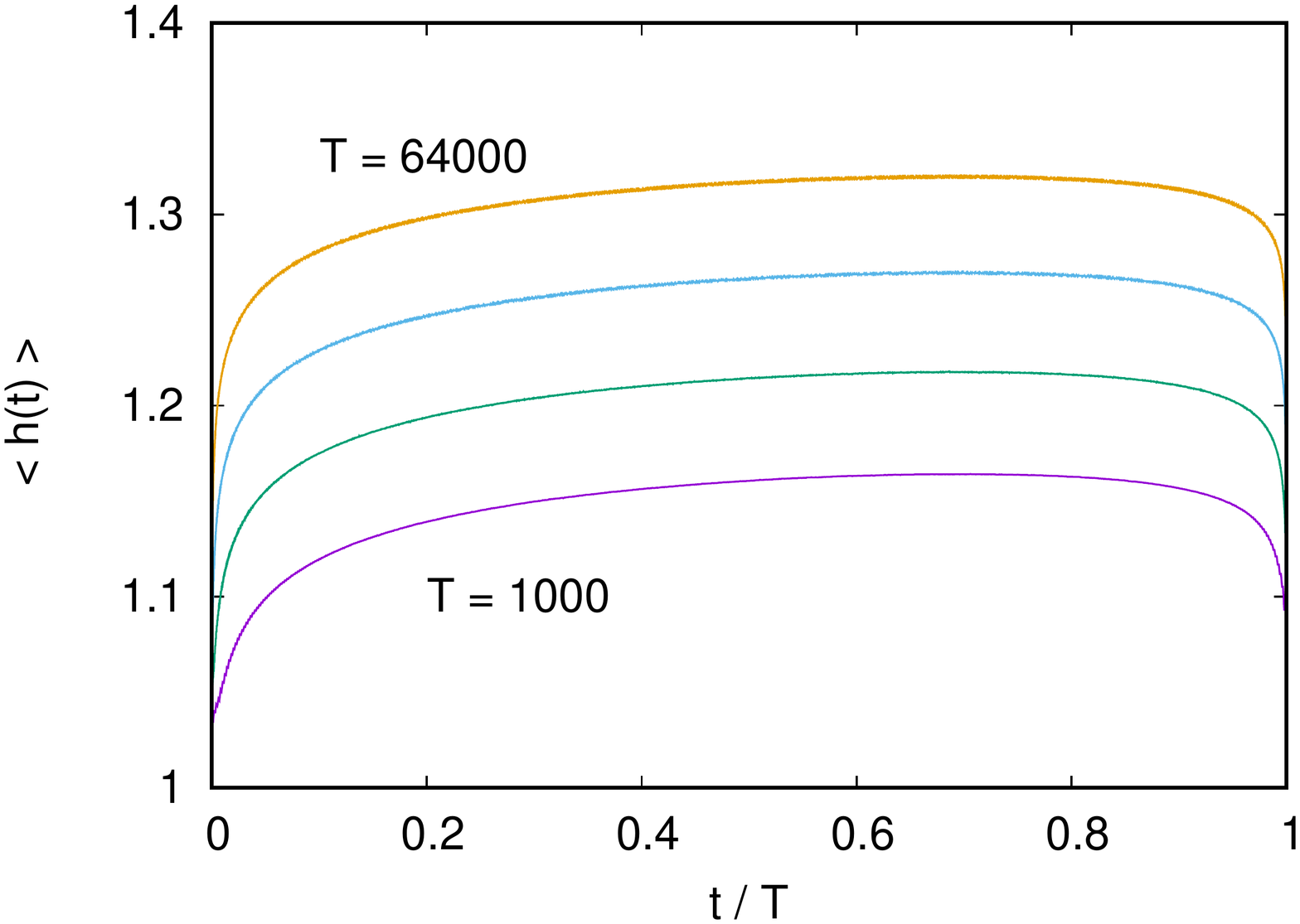}
\vglue -.6cm
\par\end{centering}
\caption{\label{figS1} (color online)  Plots of $\langle h({\bf x}(t))\rangle$, the average debris
   heights for true self-avoiding walks of total length $T$ at the position of the walker at time $t$, 
   plotted against $t/T$.
   The curves are for $T=1000, 4000, 16000$, and $64000$. The inverse temperature was $\beta=\infty$.}
\end{figure}

It is known that self-repelling walks in $d=2$ are slightly swollen. Our results on reversibility 
suggest that the effective self-repulsion is stronger at later times than at early times. On the 
other hand, we are primarily interested in the swelling of very long walks. Thus it would seem 
obvious that we should have faster convergence to the asymptotic behavior if we do not consider
the behavior of entire walks of length $T$, but the last $T$ steps of walks of total length $\gg T$.
Indeed, the first part of a very long walk is always atypical in being on a flat debris landscape.
On the other hand, if -- as we show in the main text -- the local roughness of the debris landscape
becomes statistically stationary, then the last part is always typical. Since every long TSAW 
is always part of a much longer walk (TSAWs never die, in contrast to ordinary SAWs), this means
that every TSAW is less typical for the asymptotic behavior than its last part.

Thus our strategy to study the large-$T$ behavior of the r.m.s. end-to-end distance
\be
   R^2(T) = \langle ({\bf x}(T)-{\bf x}(0))^2\rangle
\ee
should be to simulate very long walks, of total length $\gg T$, and study
\be
   \tilde{R}^2(T) = \langle ({\bf x}(t+T)-{\bf x}(t))^2\rangle
\ee
for $t\gg T$. To do so, we use a large but finite lattice of size $L\times L$ with periodic 
(more precisely: helical) boundary conditions. In order to avoid finite size corrections we 
should consider only $T$ values much smaller that $L^2$. On this lattice, a single very
long walk ($> 10^{12}$ steps) is simulated. The last $T_{\rm max} = 2^k$ positions are stored, at 
any time, and at fixed intervals $({\bf x}(t+T)-{\bf x}(t))^2$ is calculated for 
$T=1,2,4\ldots T_{\rm max}$, after dismissing a transient. 

\begin{figure}
\begin{centering}
\includegraphics[scale=0.30]{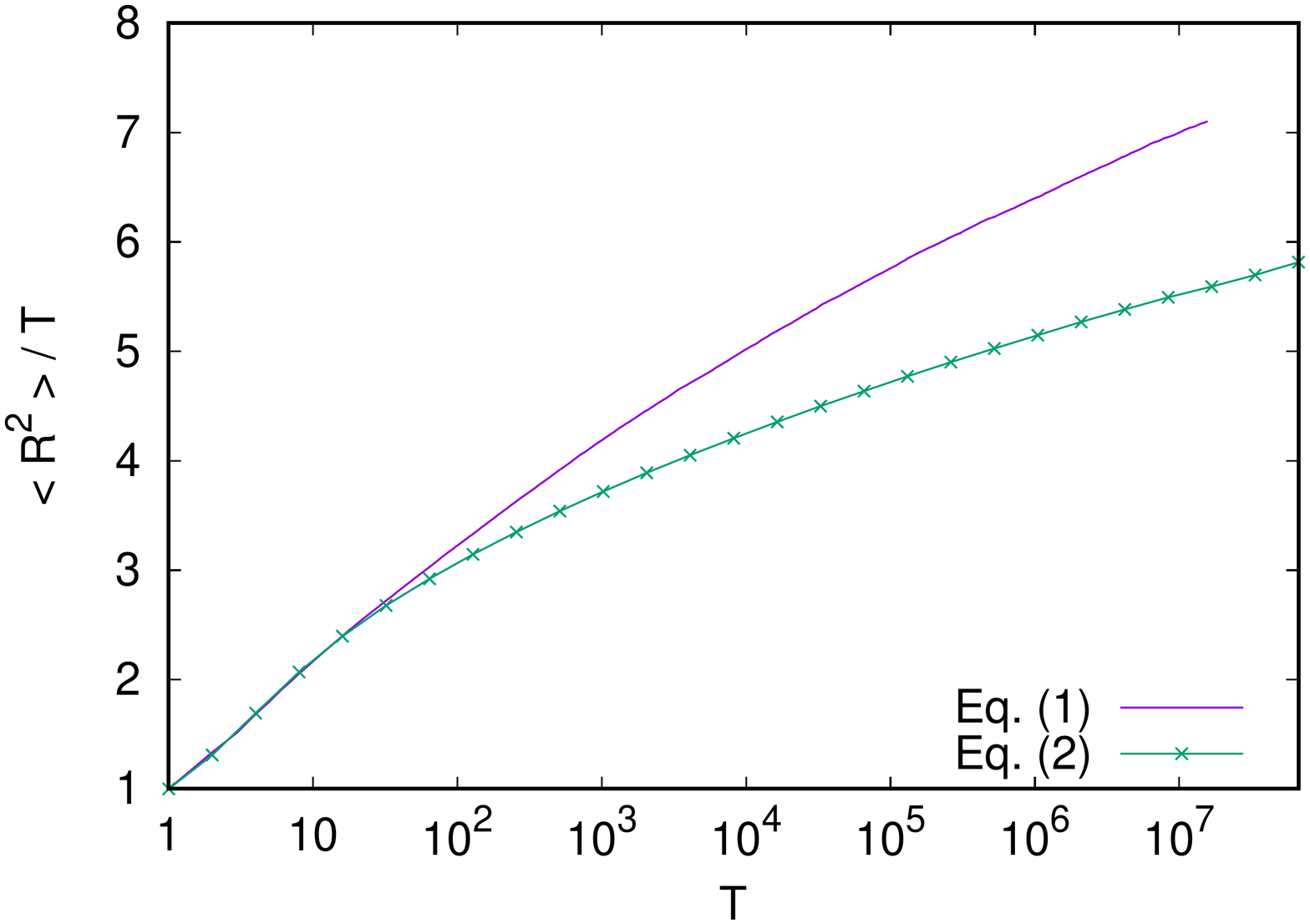}
\vglue -.6cm
\par\end{centering}
\caption{\label{figS2} (color online)  Log-linear plots of r.m.s. end-to-end distances of TSAWs of 
   length $T$, divided by $T$. In the upper curve we show this for the first $T$ steps (Eq.~(S1)), 
   while the lower curve shows data for the last parts of very long walks (Eq.~(S2)). For both
   curves, statistical errors are comparable to the line thickness. }
\end{figure}
 
\begin{figure}
\begin{centering}
\includegraphics[scale=0.30]{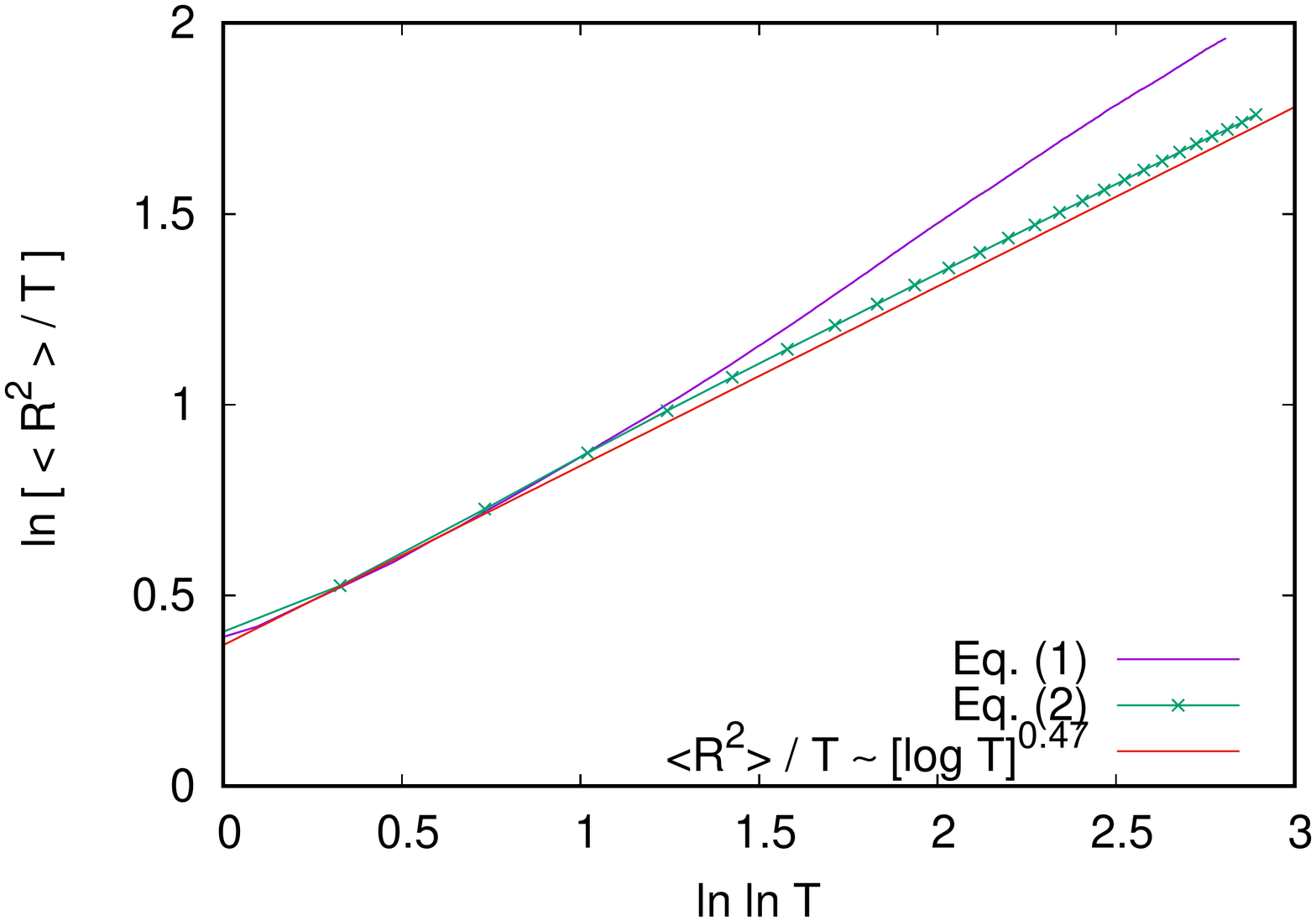}
\vglue -.6cm
\par\end{centering}
\caption{\label{figS3} (color online) Same data as in the previous figure, but plotted as 
   $\ln [ R^2/T]$ against $\ln \ln T$. A scaling law $R^2/T \sim [\ln T]^\alpha$ would give a
   straight line with slope $\alpha$. The data give $\alpha = 0.47(2)$ for the lower curve, and 
   no convincing scaling for the upper.}
\end{figure}

In Fig.~S2 we compare estimates of $R^2/T$ obtained with Eqs.(S1) and (S2) with each other. In these 
data we checked very carefully that finite size corrections are negligible. In view of the claim
that self-repulsion effects should be stronger at late times it seems strange that the estimates 
obtained from Eq.(S1) are larger than those obtained from Eq.(S2). But this is actually easily 
explained. At $t=0$, walks start on a flat landscape. At any $t \gg L^2$, in contrast, they are
in an already rough landscape, and moreover they are more likely to be in a sink instead of being
on a hilltop. Thus, when we start measuring $R^2$, the debris put down during the subsequent steps 
makes them swollen, but the debris put down already before makes then compressed.

In order to test the scaling $R^2/T \sim [\ln T]^\alpha$, we plotted in Fig.~S3 the same data as 
$\ln [ R^2/T]$ against $\ln \ln T$. We now observe exactly what we expected: While Eq.~(S1) does 
not give any convincing scaling (for small $T$ the curve bends up, but for large $T$ it bends down),
Eq.~(S2) gives a perfect straight line with $\alpha = 0.47(2)$. Within two standard deviations 
this is compatible with $\alpha =1/2$.

\begin{figure}
\begin{centering}
\includegraphics[scale=0.30]{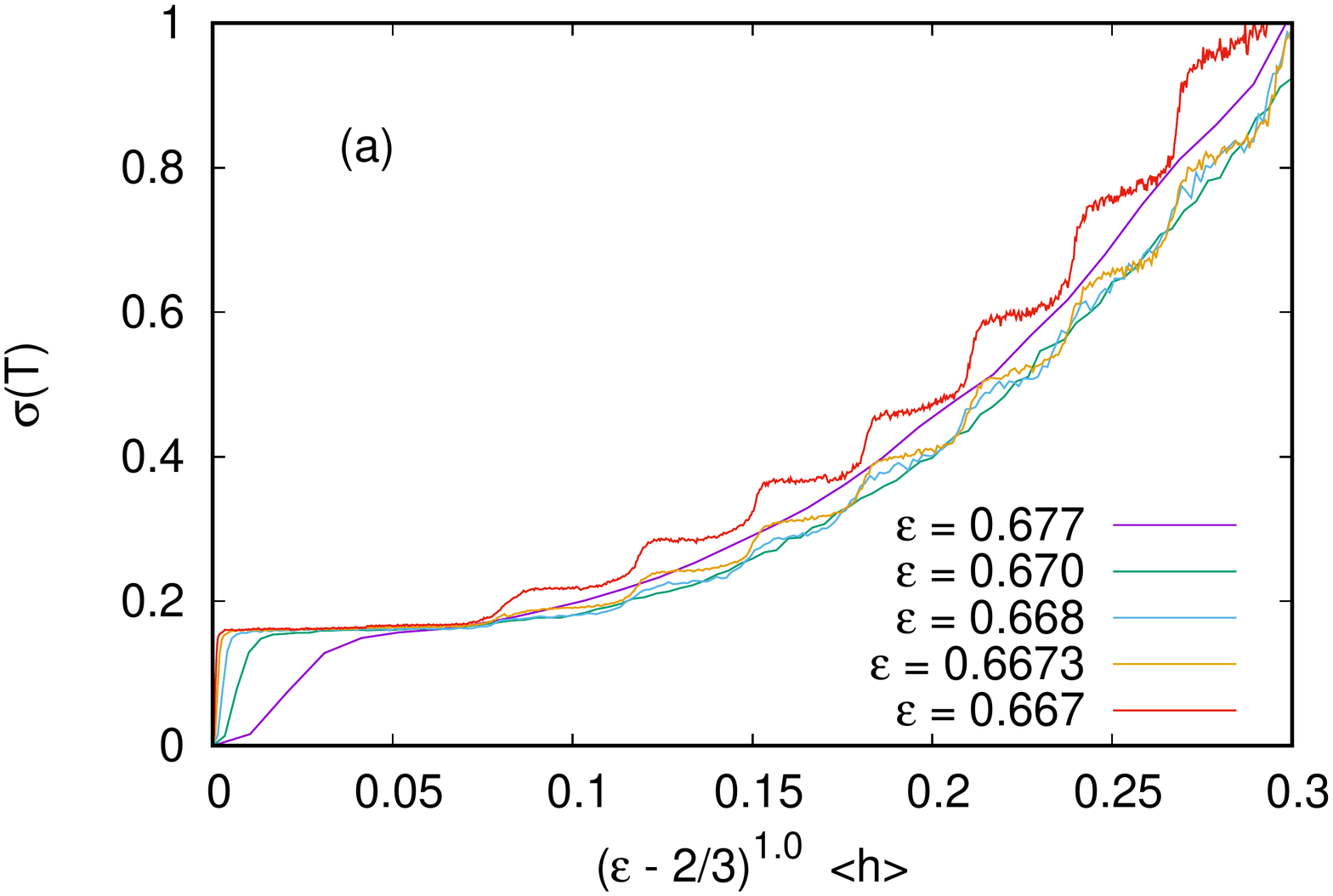}
\vglue -1.0cm
\includegraphics[scale=0.30]{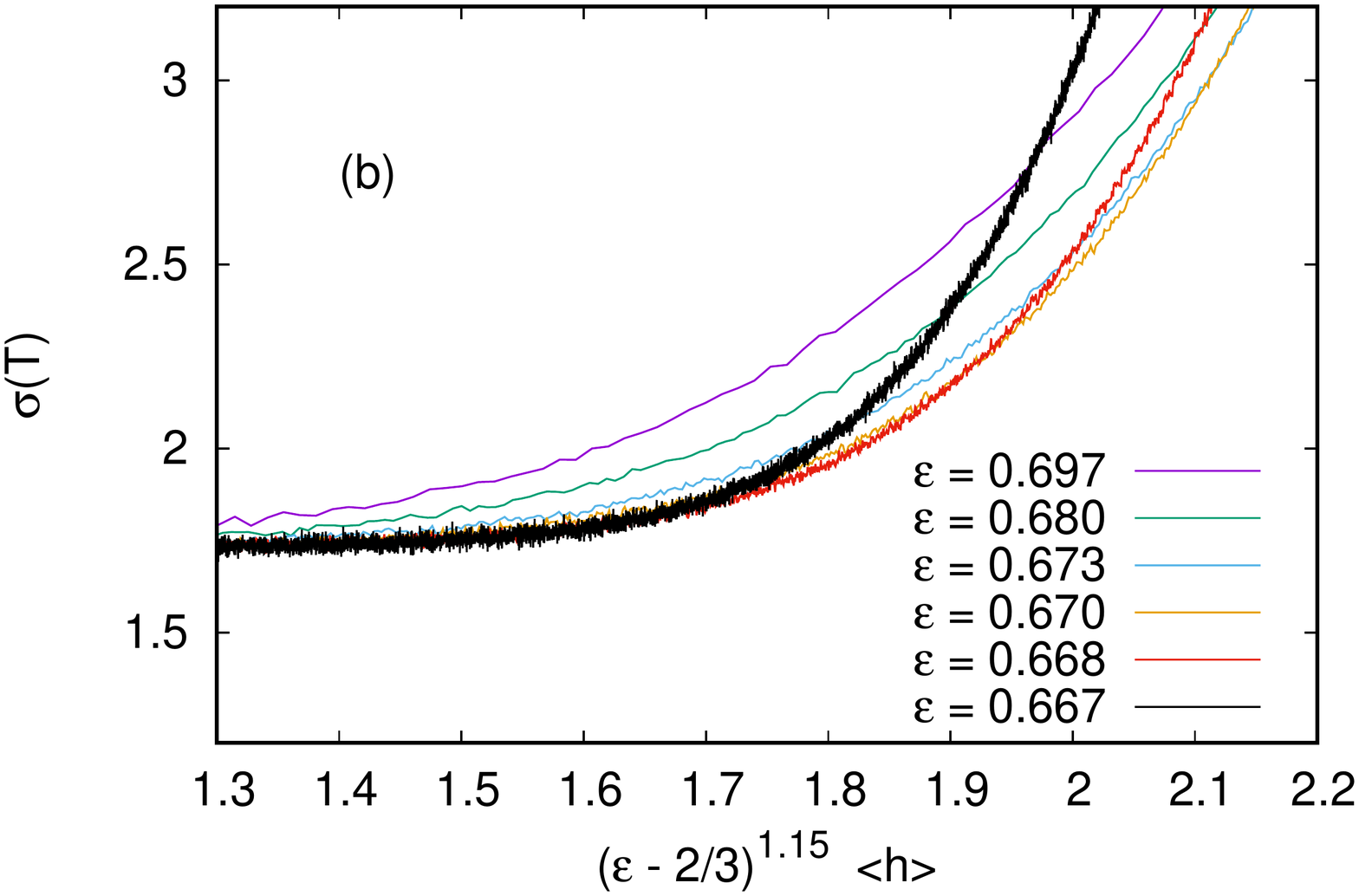}
\vglue -.6cm
\par\end{centering}
\caption{\label{figS4} (color online) Plots of $\sigma(T)$ against  $(\epsilon - 1/2)^\gamma \langle h\rangle$,
   for square lattices, and when debris is put on nearest and next-nearest neighbors. Panel (a) is for 
   $\beta=\infty$ and $\gamma = 1.0$, while (b) is for $\beta=1.0$ and $\gamma = 1.15$.}
\end{figure}

\section{Debris is also put on next-nearest neighbors}

Plots analogous to Fig.5 of the main text, but for the case when debris is put on all eight nearest and
next-nearest neighbors (with equal amounts on each) are shown in Fig.~S4. Details are as in 
Fig.5, except that now $\gamma=1.15$. A most remarkable feature in panel (a) are the steps that develop
as $\epsilon \to 2/3$. This is a further case where $\beta=\infty$ shows structures that are not seen 
for finite $\beta$.

\begin{figure}
\begin{centering}
\includegraphics[scale=0.30]{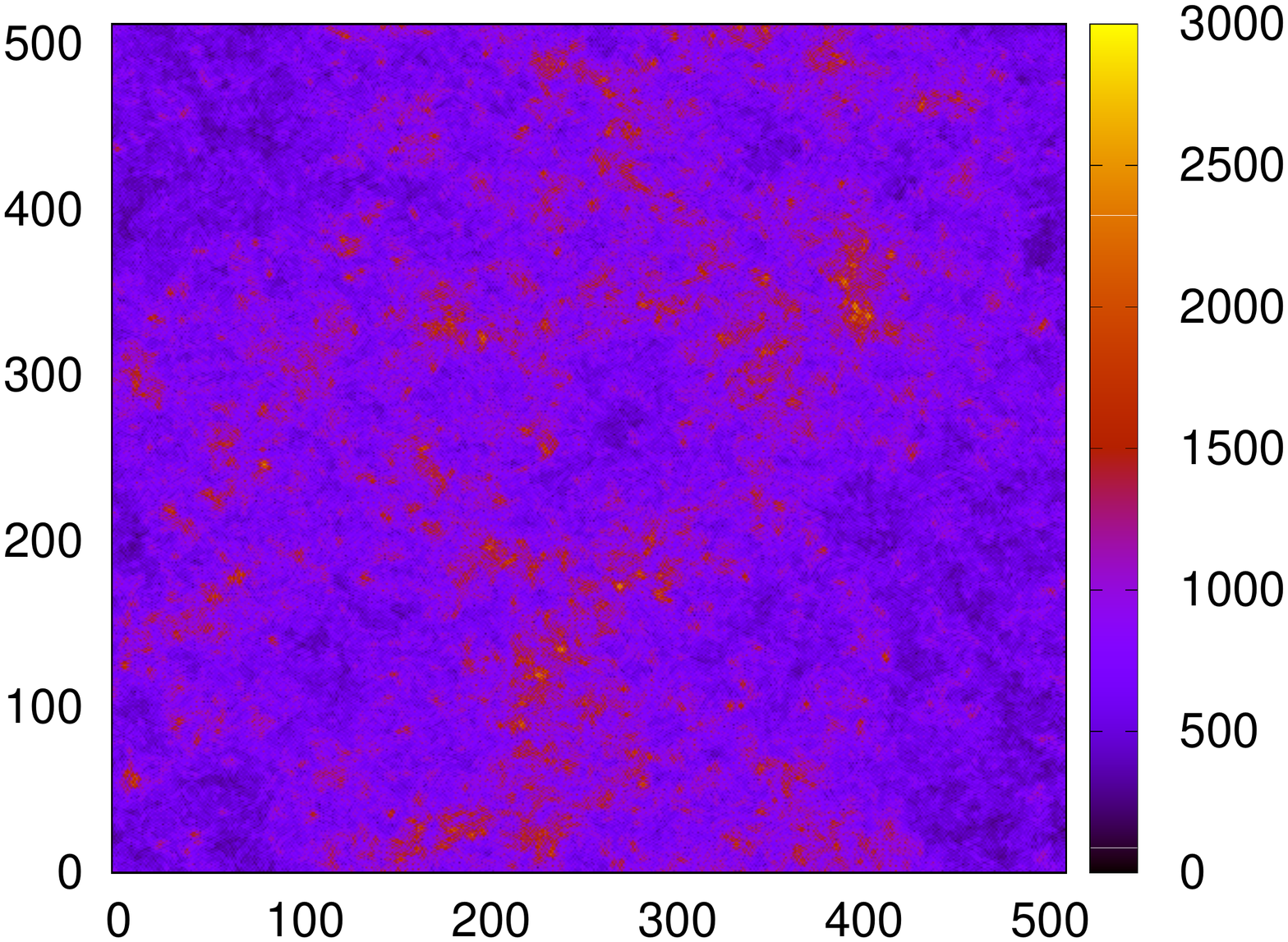}
\vglue -1.0cm
\includegraphics[scale=0.30]{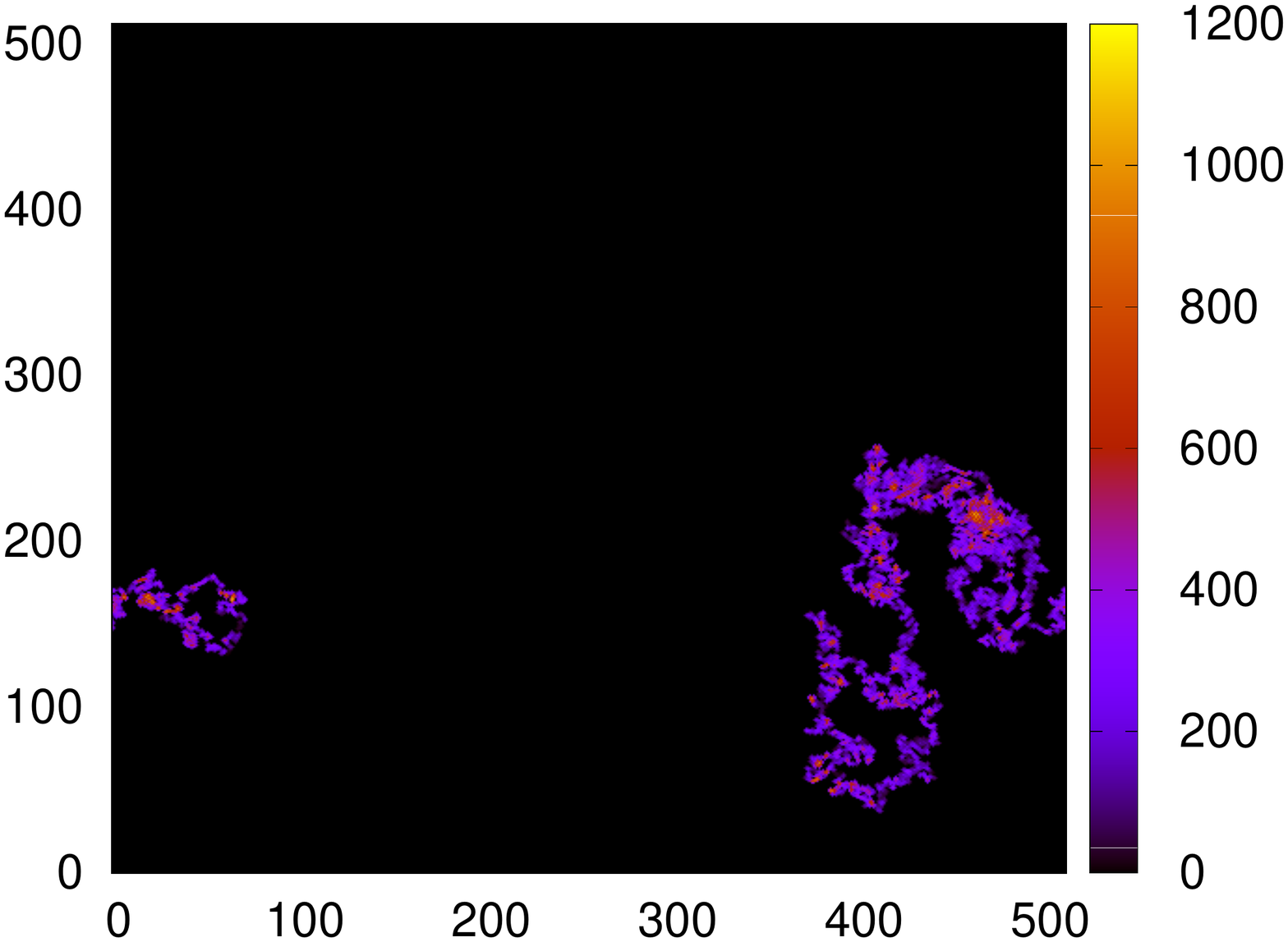}
\vglue -.6cm
\par\end{centering}
\caption{\label{figS5} (color online) Upper panel: Debris landscape for $\beta=3.0, \epsilon=0.8$, and 
   $\langle h\rangle=800$. It is very rough and without any obvious structure. Lower panel:
   height difference between this landscape and the one at a later time, when $\langle h\rangle$ has 
   increased by 10 units. Although a free random walk would have covered the entire lattice during this 
   additional time, due to the landscape roughness it stayed trapped in a very narrow region.}
\end{figure}

\section{Self-trapped walks}

In Fig.~S5 we show the debris landscape for typical parameter values where the walk is self-trapped 
(upper panel). In the lower panel we show the height difference between this landscape and the one at a 
later time, when the average height has increased by 10 units. We see that the landscape is very rough 
indeed, and that the walks have not ventured of a rather narrow region during the successive time steps, 
i.e. they really are trapped during rather long periods.

The entropy (in bits) per step at time $t$ of a random walk (or, more generally, of any stochastic 
process) is simply $-\sum_i p_{t,i} \log_2 p_{t,i}$, where the sum runs over all moves possible at $t$ 
and $p_{t,i}$ is the probability of the $i$-th possible move. For walks on a square lattice with 
$\beta=\infty$ it is simply $\log_2 m_t$, where $m_t$ is the number of neighbors with the smallest 
debris height at time $t$. Notice that we do not need any assumption on stationarity or any Markov
property.

\begin{figure}
\begin{centering}
\includegraphics[scale=0.30]{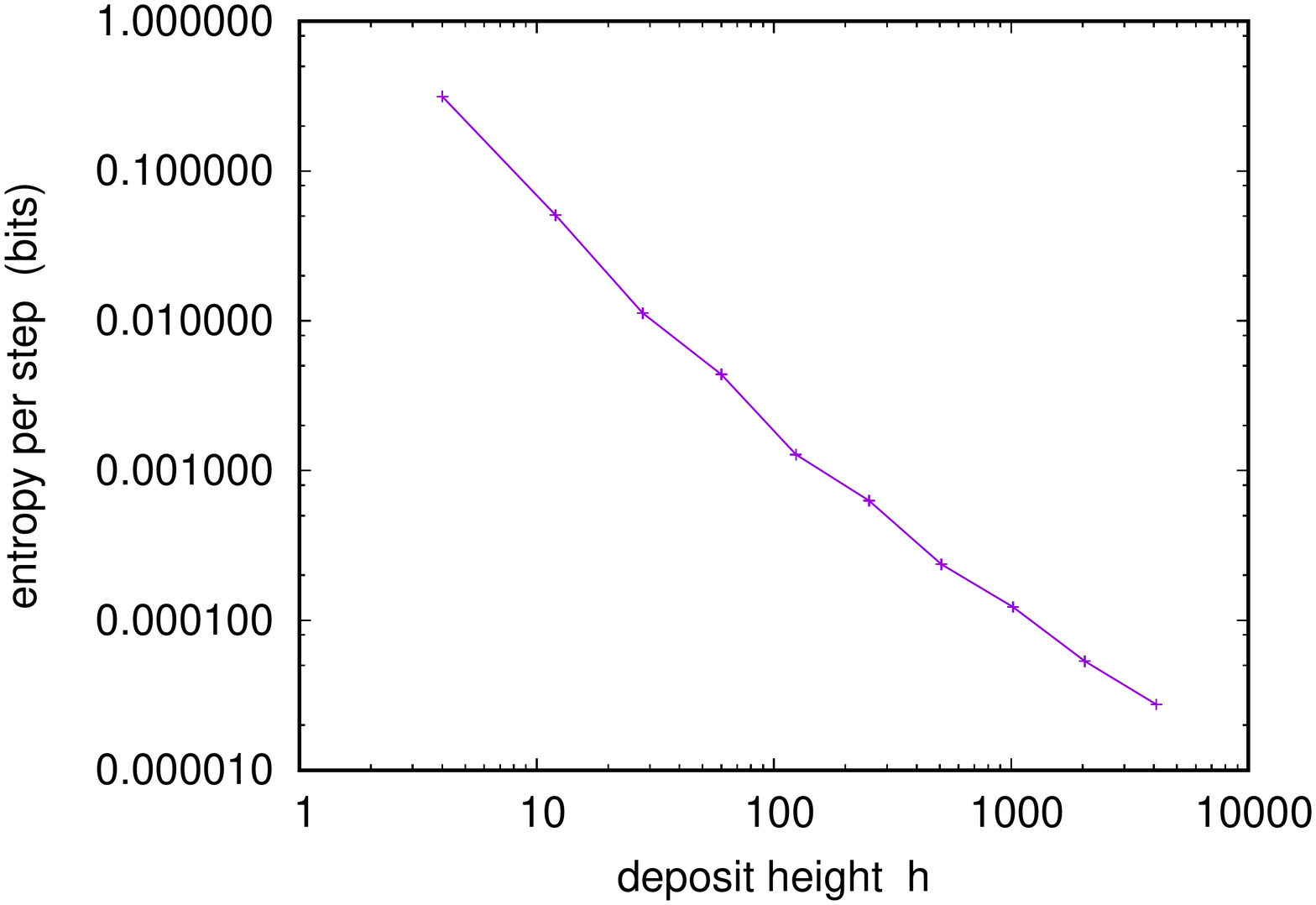}
\vglue -.6cm
\par\end{centering}
\caption{\label{figS6} (color online) Entropy per step (in bits), averaged over time intervals 
   that correspond to heights increasing from the value $h$ plotted on the x-axis to the next value.
   In this plot the square lattice is used with $L = 16384, \beta=\infty$ and $\epsilon=0.7$}
\end{figure}

In Fig.~S6 we show results for $\beta=\infty$ and $\epsilon = 0.7$. The entropies here are averaged over 
increasingly long bins in $t$ (or, equivalently, in $h=t/L^2$). The horizontal axis shows the heights
at which the averaging started. The extremely small entropies at large deposit heights correspond to walks
that most of the time just go back and forth between two neighboring sites.

\section{Pattern formation}

\begin{figure}
\begin{centering}
\includegraphics[scale=0.30]{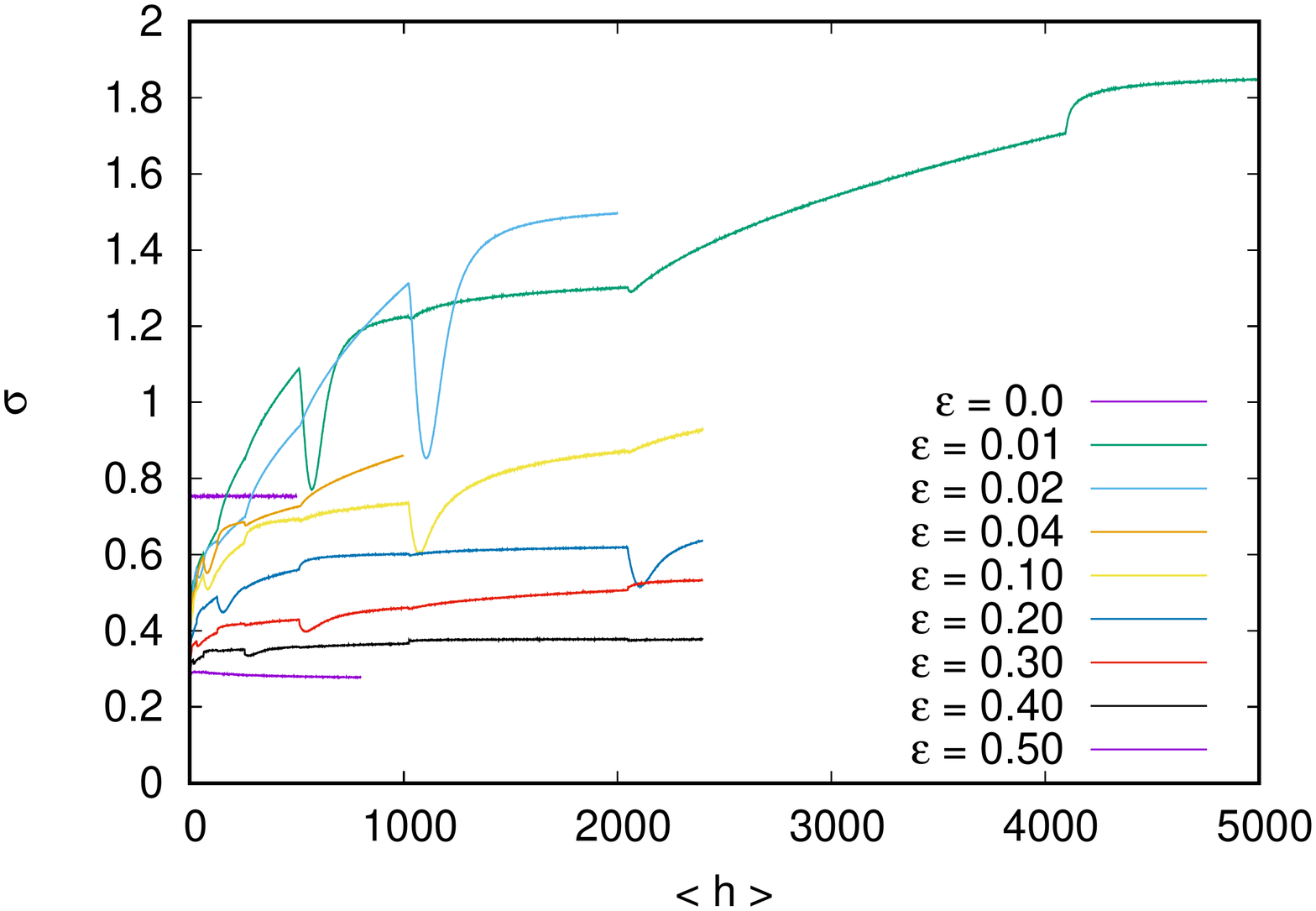}
\vglue -.6cm
\par\end{centering}
\caption{\label{figS7} (color online) Height variances plotted against average height for fixed $\epsilon$.
   All curves are for $L=512$ and $\beta=\infty$, and the values of $\epsilon$ are such that the walks
   are superdiffusive.}
\end{figure}

\begin{figure}
\begin{centering}
\includegraphics[scale=0.30]{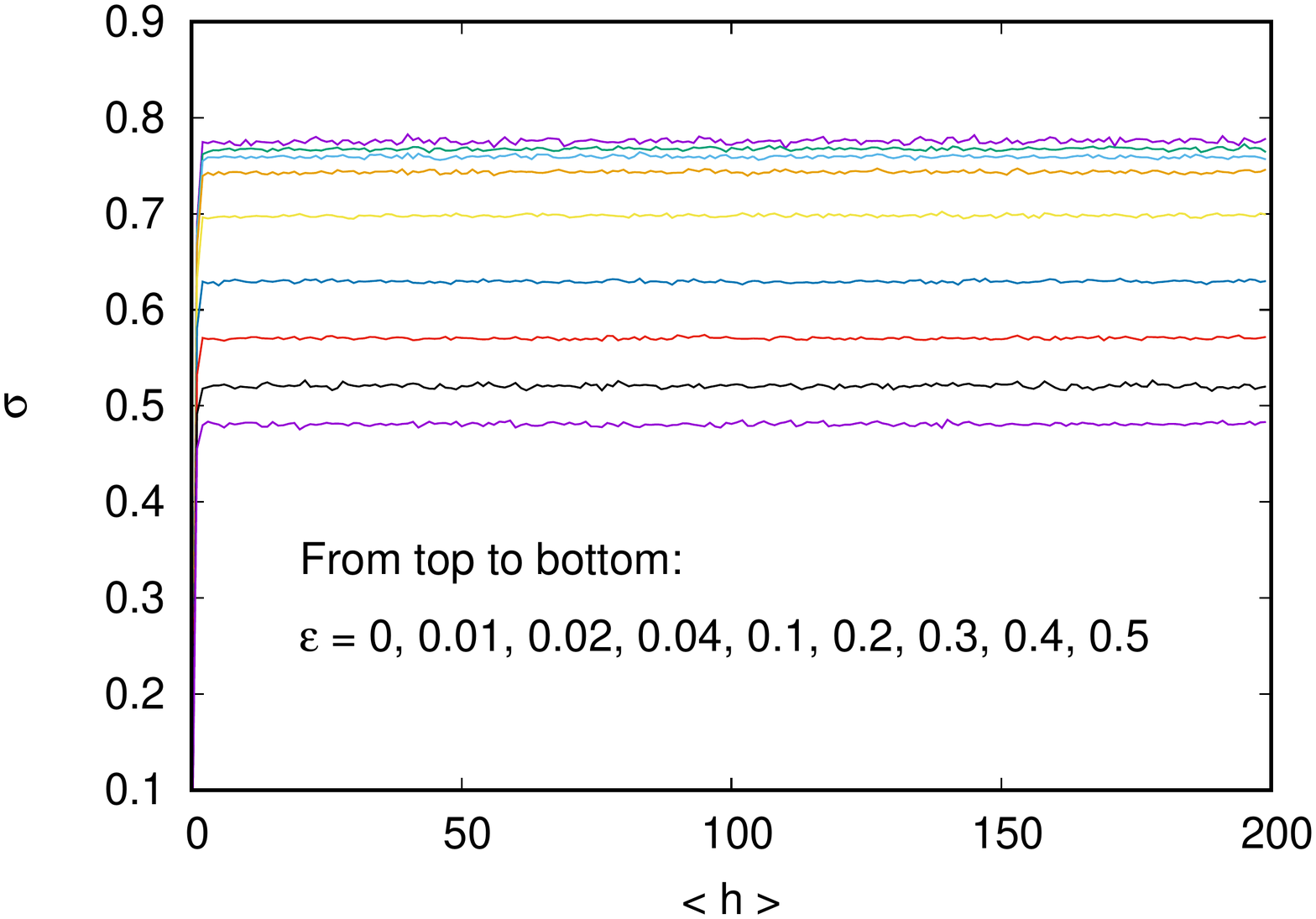}
\vglue -.6cm
\par\end{centering}
\caption{\label{figS8} (color online) Same as the previous figure, but for $\beta=5.0$.}
\end{figure}

\begin{figure}
\begin{centering}
\includegraphics[scale=0.30]{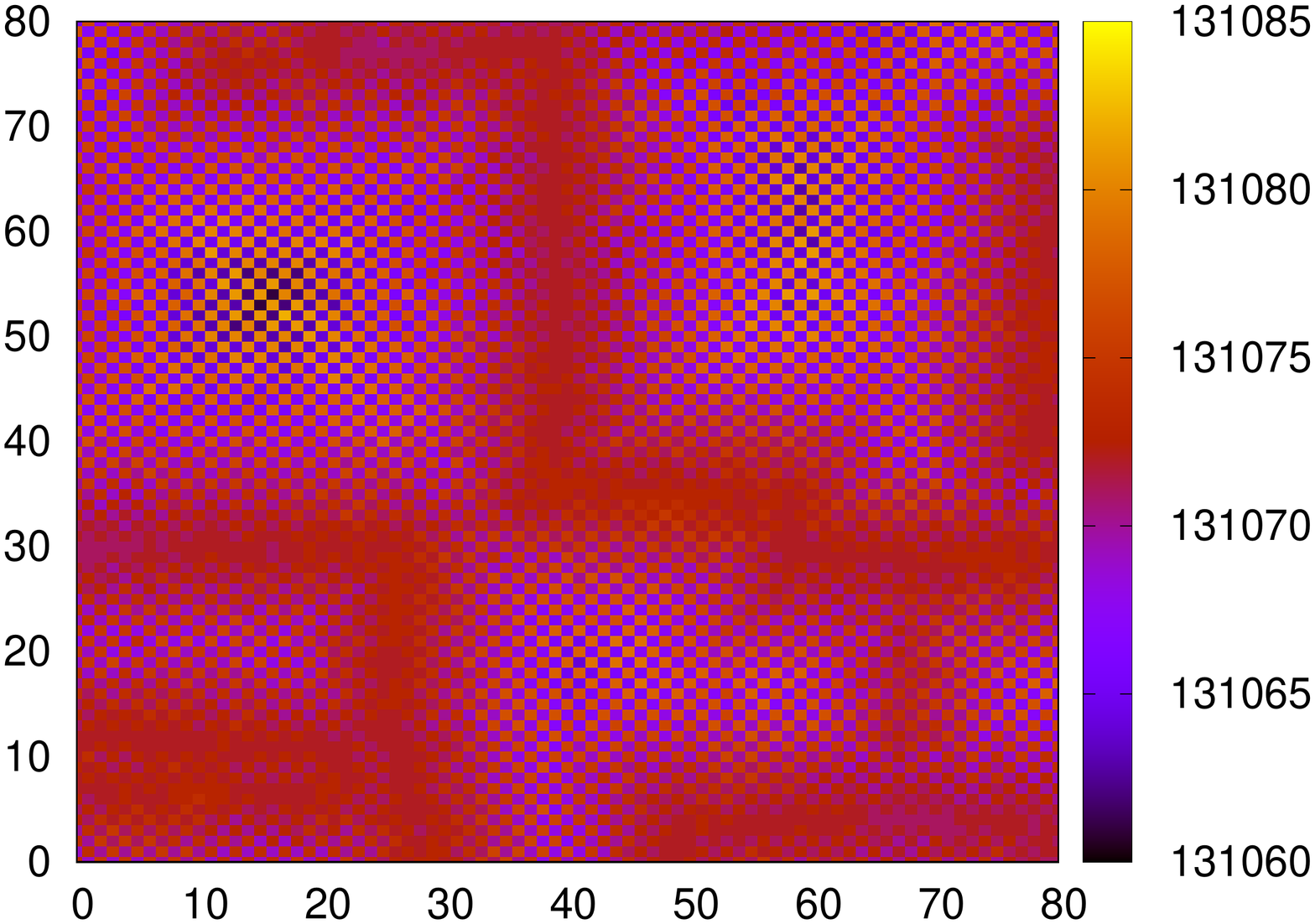}
\vglue -.9cm
\includegraphics[scale=0.30]{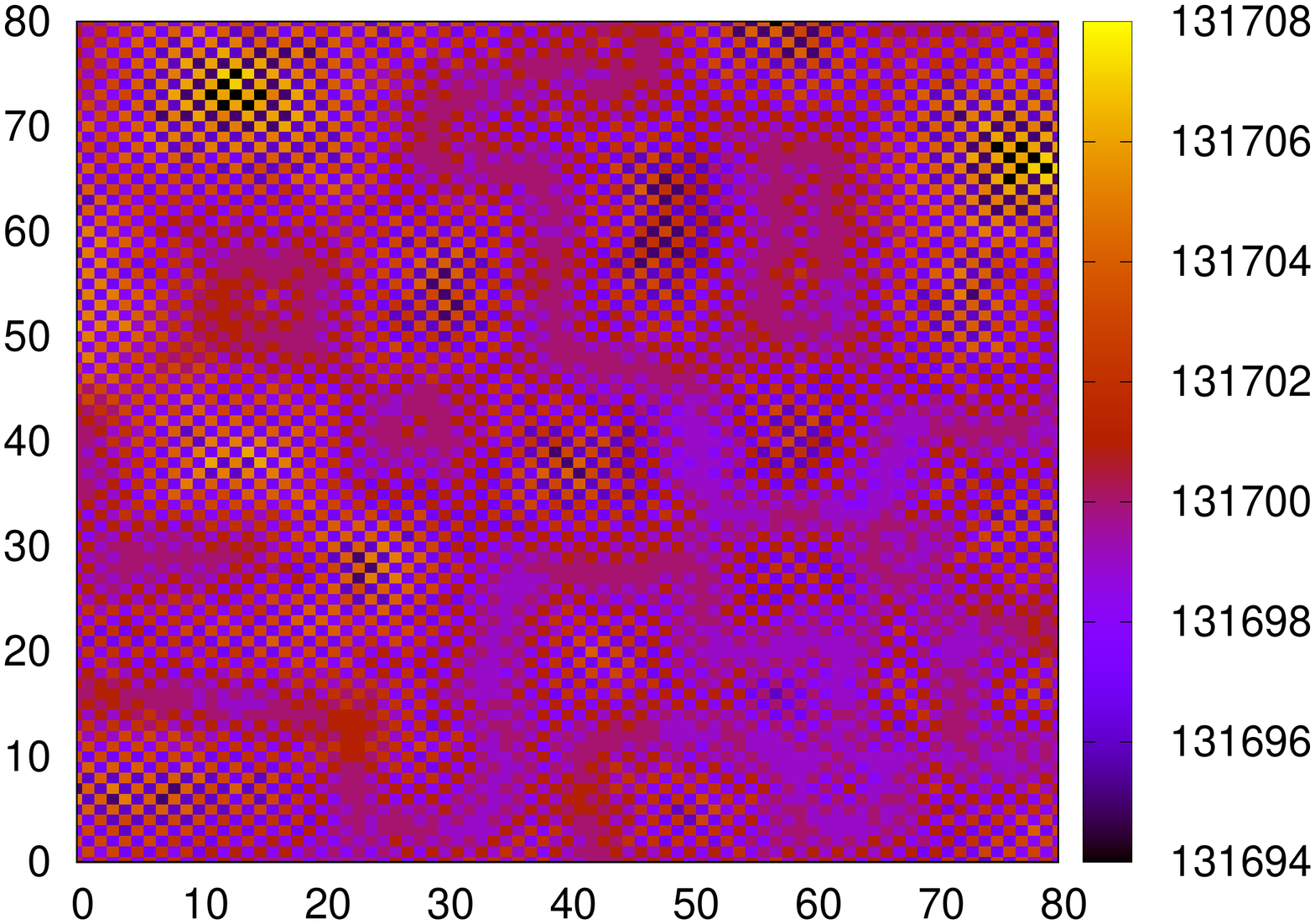}
\vglue -.6cm
\par\end{centering}
\caption{\label{figS9} (color online) Part of debris landscapes for $\beta=0,\epsilon=0.0012, L=256$. The 
   upper panel is for $\langle h \rangle = 131072$ (where $\sigma = 10.06$), while the lower one is for 
   $\langle h \rangle = 131700$ (where $\sigma = 4.87$). Both snapshots were taken during the same run. 
   Notice that regions with very strong checkerboard patterns in the former have weak such patterns in 
   the latter.}
\end{figure}

We have already mentioned several cases (see Figs.~2 and 3 in the main text, and Fig.~S4a) where modified 
self-repelling walks with $\beta=\infty$ are more complex or irregular than the walks with $\beta<\infty$. 
Remember that walks {\it always} go
to the neighbor sites with lowest $h$ when $\beta=\infty$, while there is only a finite bias when 
$\beta<\infty$. Here we discuss another striking instance where the behavior is more complex for 
$\beta=\infty$ than for $\beta<\infty$.

\begin{figure}
\begin{centering}
\includegraphics[scale=0.30]{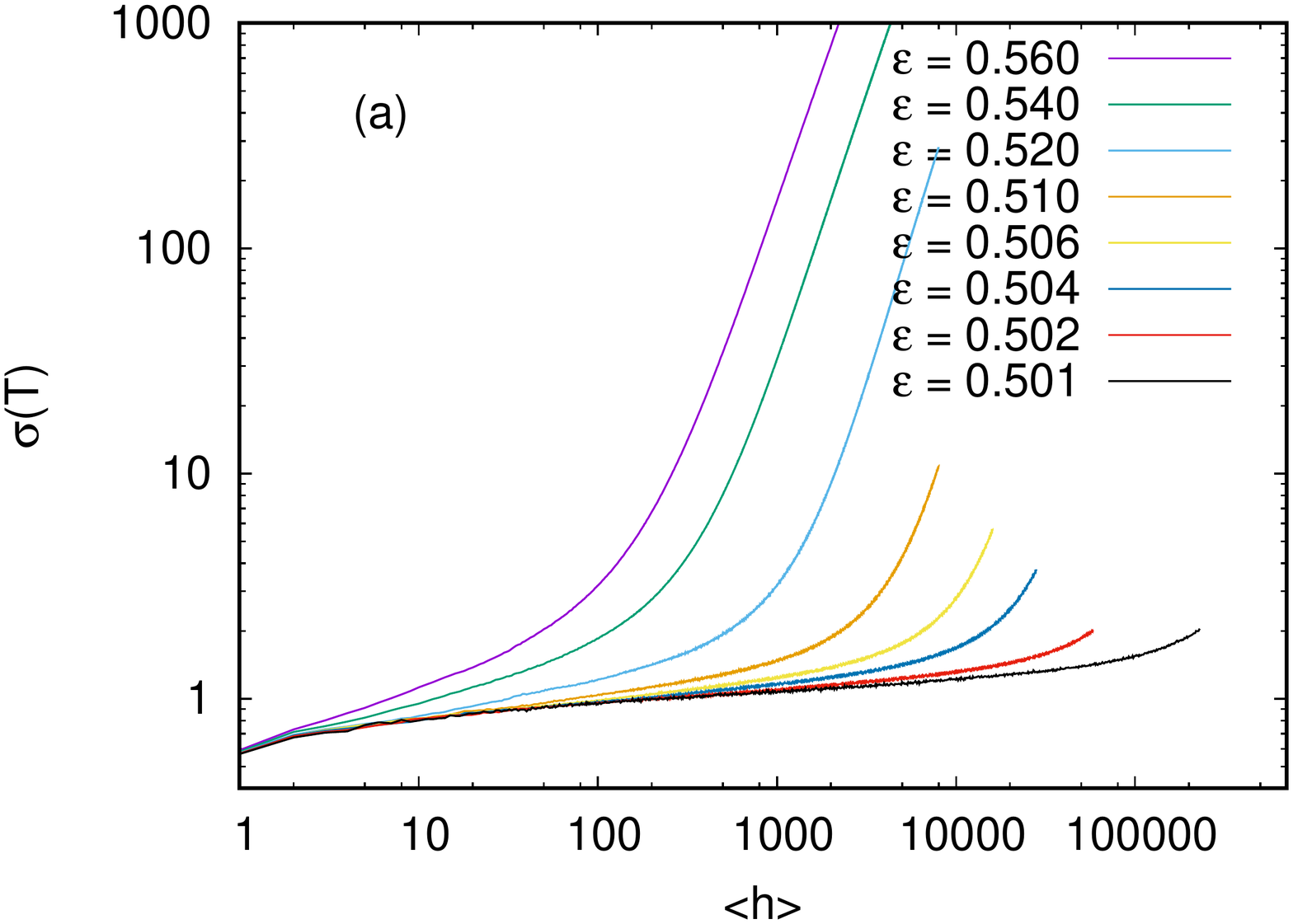}
\vglue -1.0cm
\includegraphics[scale=0.30]{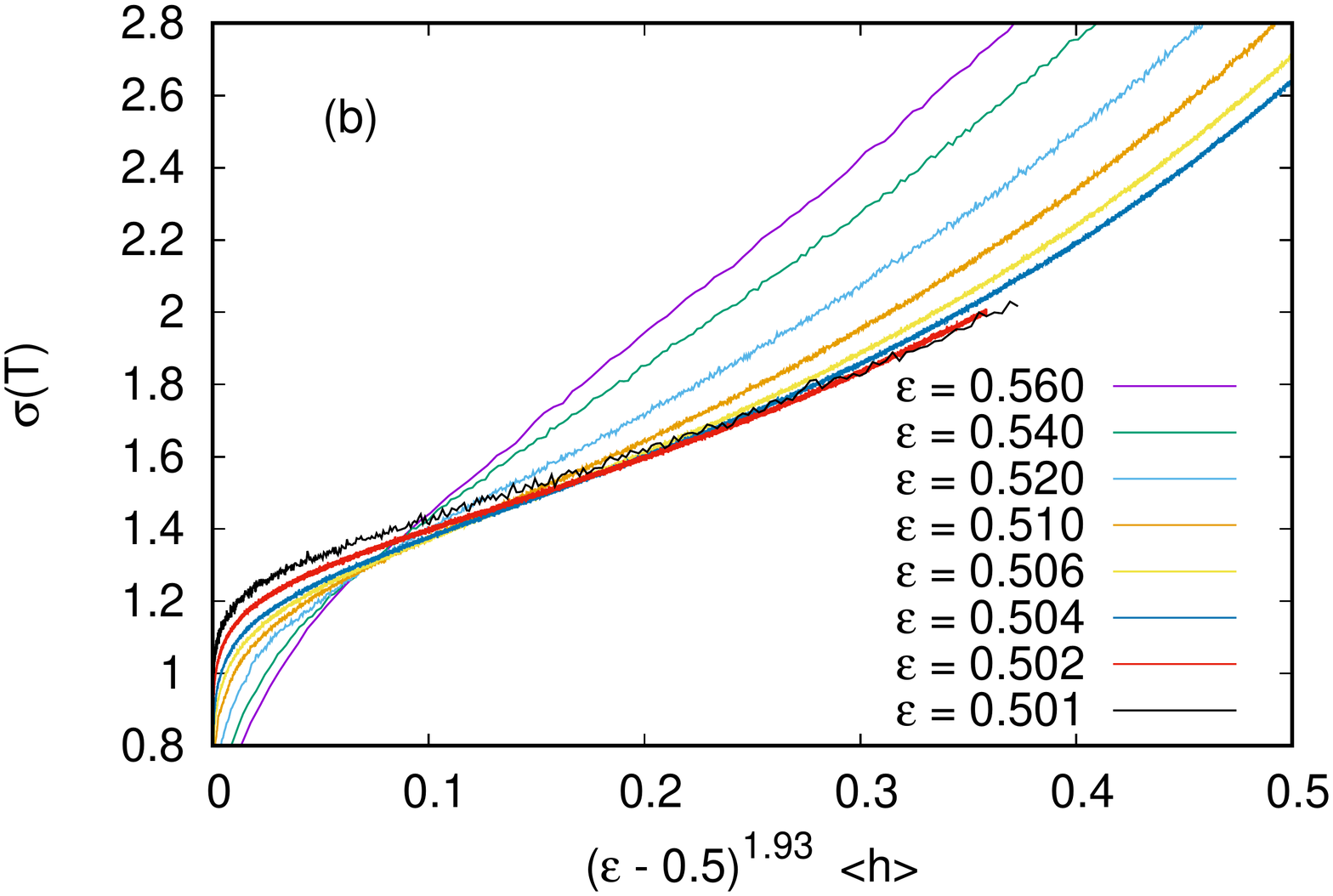}
\vglue -1.0cm
\includegraphics[scale=0.30]{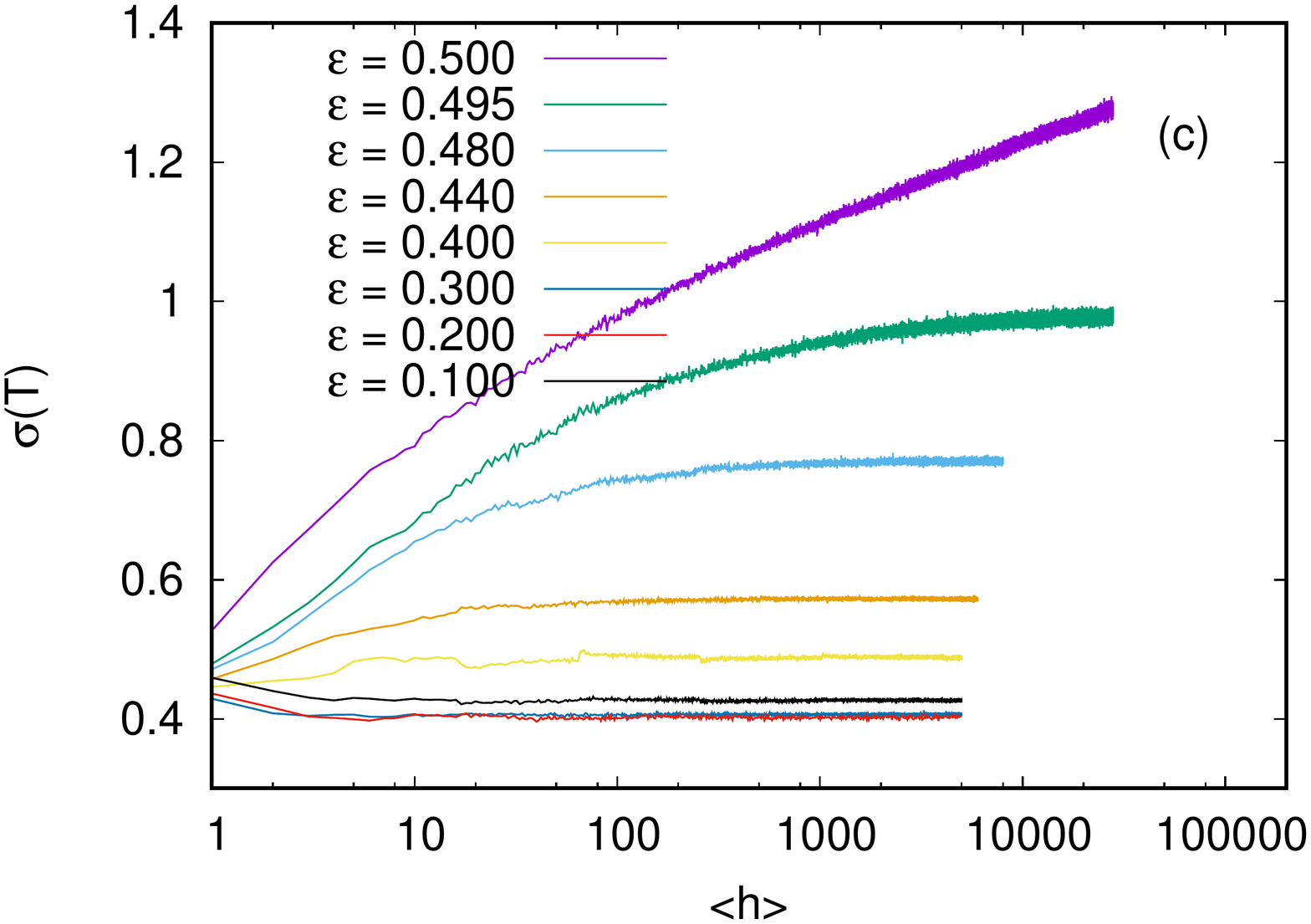}
\vglue -.6cm
\par\end{centering}
\caption{\label{figS10} (color online) Plots of $\sigma(T)$ for the model in which debris is put onlu on 
   a single randomly chosen eighbor. Panel (a): Values for $\epsilon > 1/2$ plotted against 
   $\langle h \rangle$. (b): Values for $\epsilon > 1/2$ plotted against $(\epsilon - 1/2)^1.93
   \langle h \rangle$. (c): Values for $\epsilon \leq 1/2$ plotted against $\langle h \rangle$.}
\end{figure}

More precisely, let us consider variances of $h$ at finite times on square lattices, with $\beta=\infty$ 
and $0 < \epsilon < \epsilon_c$. In this case the height variances stay finite. For $\epsilon = 0$ they
rapidly approach a constant, and for $\epsilon = \epsilon_c$ they slowly and monotonically decrease after
a fast initial rise (see Fig.~2). But as seen from Fig.~S7, they vary strongly and seemingly irregularly 
when $0 < \epsilon < \epsilon_c$. We have the following comments:
\begin{itemize}
\item All structures seen in Fig.~S7 are statistically significant. Statistical errors are comparable 
to the line width.
\item The curves do not change when the helical b.c. are replaced by periodic ones.
\item Although the curves are shifted vertically when $L$ is changed, the shapes of the curves are 
independent of the system size.
\item Although the dependence on $\epsilon$ seems very unsystematic, all curves have one feature in
common: They show slightly rounded kinks when $\langle h\rangle$ is a power of 2. Some of these kinks are 
up-turns, at others the curves go down. It seems that these kinks become sharper with increasing 
$\langle h\rangle$ and for increasing $L$. If this is true, then each kink would be a phase transition.
\item This pattern disappears entirely for finite $\beta$, see Fig.~S8.
\item The height landscapes show large patches where the heights form checkerboard-like patterns, see 
Fig.~S9. At the kinks, these patches are strongly reduced or increased. This might also explain why similar 
patterns are not seen on triangular lattices.
\end{itemize}

While the formation of checkerboard-like patterns might not be too surprising, it seems entirely unclear 
why the kinks are located at heights that are a power of 2, and why this is true for all $\epsilon$.

\section{Stochastic debris spreading on neighbors}

It seems that the structures seen in the last section are due to the lack of stochasticity when $\beta=\infty$.
To verify that this is indeed the case, and that any additional source of stochasticity would destroy them,
we finally simulated a version where the debris that is given to neighbors is not distributed 
uniformly among them, but randomly. More precisely, we now put a fraction $1-\epsilon$ of the debris onto
the present site of the walker, while the remaining fraction $\epsilon$ is put on a single randomly chosen
neighbor. 

Results obtained in this way, for $\beta=\infty$, are shown in Fig. S10. We see that:
\begin{itemize}
\item The critical value of $\epsilon$ is again $1/2$, as for the case where debris is uniformly 
given to all neighbors.
\item There is no sign of any structure or irregularity, in conrast to the case where debris is 
  distributed uniformly. This is true both for $\epsilon>\epsilon_c$ and for $\epsilon<\epsilon_c$.
\item The exponent $\gamma$ is again close to 2.
\item For $\epsilon=\epsilon_c$ the variance now increases slowly, compatible with a logarithmic
increase.
\end{itemize}
 
Finally, we have also simulated two more variants, both for $\beta=\infty$. In the first we use again the 
square lattice with debris put on the next neighbors, but now 2 different randomly chosen neighbors
receive an amount $\epsilon/2$ each. Again we find (data not shown) $\epsilon_c=1/2$ and $\gamma \approx 2$,
and no sign of any irregularities as in Figs.~3, S4a, or S7.

In the second we also use the square lattice, but we put an amount $\epsilon$ of debris onto one randomly
chosen site among the 8 next and next-to-next neighbors. As in the case when debris is spread uniformly 
among all 8 next and next-to-next neighbors, we have again $\epsilon_c=2/3$, and again $\gamma \approx 1$,

In summary, it seems thus that $\epsilon_c$ and $\gamma$ are independent of $\beta$ and of the way {\it how} 
debris is spread onto neighbors, while they do depend on the lattice type and on the set of neighbors that 
receive debris.